\def\gore{E.\,CHMIELEWSKA, M.\,TOMCZAK: Hot and cool plasma ejections}

\documentclass{ceab}   

\usepackage{epsfig}     
\usepackage{graphicx}   

\usepackage{ceabbib}     
\usepackage[T1]{fontenc}

\begin{document}

\title{Hot and cool plasma ejections in the solar corona}

\author{\normalsize  E.\,CHMIELEWSKA, M.\,TOMCZAK \vspace{2mm} \\
        \it Astronomical Institute, University of Wroc{\l }aw, \\
        \it  ul.\,Kopernika 11, PL-51-622 Wroc{\l }aw, Poland}

\maketitle

\begin{abstract}
We present the results of the first attempt of statistical research
devoted to the association between X-ray Plasma Ejections (XPEs) and
prominences. For this aim, we compared contents of a catalogue of
XPEs, observed by the Soft X-ray Telescope onboard {\sl Yohkoh}, to
H$\alpha$ reports from the {\it Solar-Geophysical Data}. We found
that only less than one third of XPEs shows a low-temperature
counterpart. A modest connection between hot and cool plasma motions
in the corona is also supported by by a frequent discrepancy between
morphology and kinematics of simultaneously occurring XPEs and
prominences. We explain the poor correlation (20-30\%) between XPEs
and prominences and high correlations ($\sim$70\%) between these
phenomena and CMEs as a proof of existence of two separate
subclasses of CMEs.
\end{abstract}

\keywords{Sun: corona - X-ray Plasma Ejections (XPEs) - prominences}

\section{Introduction}

Eruptive phenomena observed above the solar surface have been
continuously attracting attention for many years. Due to their rapid
evolution and plenty of possible configurations, they are the most
spectacular events on the Sun. Moreover, eruptive phenomena play a
crucial role in shaping terrestrial space weather. The solar
magnetic field -- continuously evolving in countless episodes of
generation, reconfiguration, and decay -- is responsible for plenty
of events observed in different temporal and spatial scales in a
broad range of wavelengths. There are three main classes of events:
prominences, coronal mass ejections (CMEs), and X-ray plasma
ejections (XPEs).

Prominences are relatively cool and relatively dense parts of the
solar atmosphere routinely observed in the hydrogen H$\alpha$ line
in many ground-based observatories. They are known since 1860's
(Secchi, 1875). Among many phenomenological classifications
organizing a heterogeneous world of prominences (Tandberg-Hanssen,
1995), the kinematical one (de Jager, 1959) resolving quiescent and
moving events seems to be the most useful.

CMEs are sudden, large-scale expulsions of magnetized plasma
observed high in the corona and in the heliosphere. Since a launch
of the first space white-light coronagraph (Tousey, 1973), we can
detect the photospheric light scattered by electrons kept within
expanding magnetic structures. Recently operating coronagraphs
onboard the {\it SOHO} and {\it STEREO} satellites derive
continuously new examples of well-observed CMEs.

XPEs are sudden expulsions of hot magnetized plasma in the solar
corona seen in X-rays. XPEs display a wide range of macroscopic
motions showing different morphological, kinematic, and physical
conditions. They occur usually during the impulsive phase of flares.
XPEs have been systematically observed since 1991 when the {\sl
Yohkoh} satellite began operations (Shibata {\it et al.}, 1995). A
strong inhomogeneity of XPEs as a group suggests different physical
mechanisms responsible for their occurrence.

\section{Association between eruptive phenomena}

\begin{figure}[t]
\epsfig{file=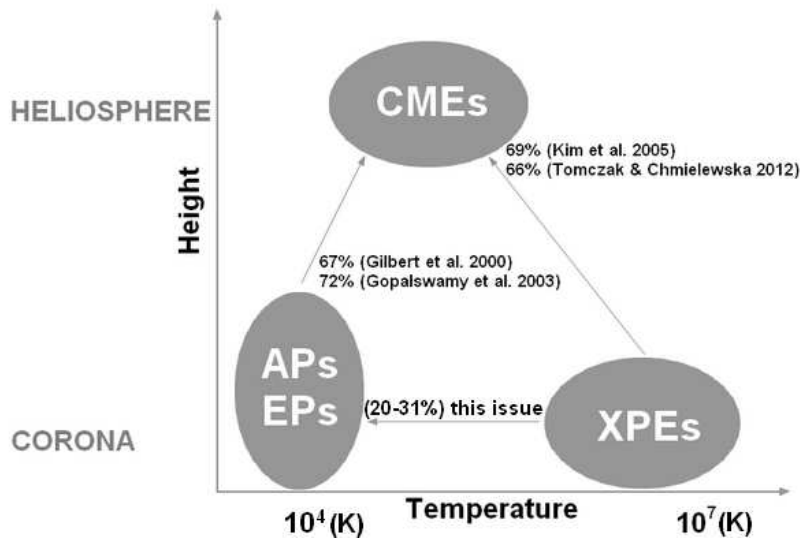,width=10.5cm} \caption{Schematic diagram
showing basic features of three major classes of eruptive phenomena
observed above the solar surface. The values written near connecting
arrows refer to levels of correlation obtained in the cited papers.}
\end{figure}

We summarize schematically basic features of solar eruptive
phenomena in a diagram temperature vs. height (Figure~1). Even this
simple illustration teaches us that simultaneously occurring events
belonging to different classes may be somehow associated one to
another, responding a reconfiguration of magnetic field in the solar
atmosphere. Therefore, statistical investigation combining lists of
events detected by particular instruments is very useful for better
understanding of this complex problem.

Observations suggest a good correlation of moving prominences with
CMEs. For the 54 these prominences observed at the Mauna Loa Solar
Observatory (MLSO) between 1996 February and 1998 June, Gilbert {\it
et al.} (2000) found 36 associated CMEs observed by the LASCO
coronagraphs onboard {\it SOHO} or by the Mark-III coronametr at the
MLSO, which gives a 67\% correlation. Particular subclasses of
prominences show different levels of the correlation: 94\% (17 from
18 events) for eruptive prominences, 70\% (7 from 19 events) for
disappearing filaments, and 46\% (12 from 26 events) for active
prominences.

Gopalswamy {\it et al.} (2003) compared a list of 186 moving
prominences, observed in the years 1996-2001 by the Nobeyama
Radioheliograph, to the {\it SOHO} LASCO CME Catalog (Gopalswamy
{\it et al.}, 2009). They found 134 associated events, i.e. a 72\%
correlation. They found also that the prominences that move radially
(a feature of eruptive prominences) show a 83\% correlation (126
from 152 events), whereas the prominences, which move transversally
(a feature of active prominences) show a 24\% correlation (8 from 34
events).

A similarly close association between XPEs and CMEs was reported.
Kim {\it et al.} (2005) compared a list of 137 XPEs, observed
between 1999 April and 2001 March by the SXT telescope (Tsuneta {\it
et al.}, 1991) onboard {\it Yohkoh}, to the {\it SOHO} LASCO CME
Catalog. They found 95 associated events, i.e. a 69\% correlation.
They found also a morphological dependence of the CME association: a
100\% correlation (for 11 events) for jet-type XPEs, a 77\%
correlation (46 from 60 events) for loop-type XPEs, a 70\%
correlation (28 from 40 events) for spray-type XPEs, and a 28\%
correlation (5 from 18 events) for confined XPEs.

Tomczak \& Chmielewska (2012) checked up the full list of XPEs
observed by the SXT/{\it Yohkoh}, when the LASCO coronagraphs were
operational, and found that 182 XPEs from 275 were associated with
CMEs. It gives a 66\% correlation. For different subclasses of XPEs
established morphologically and kinematically they found a broad
range of correlation between 44\% and 88\%.

An association between XPEs and prominences has been poorly
investigated until now. There are only a few papers, in which this
relation was monitored for singular events (Gopalswamy {\it et al.},
1997; Mari$\check{\rm{c}}$i$\acute{\rm{c}}$ {\it et al.}, 2004;
Vr$\check{\rm{s}}$nak {\it et al.}, 2004; Ohyama \& Shibata, 2008;
Kim {\it et al.}, 2009). These works show that relation between hot
and cool plasma ejections can be complex and cannot be considered as
a substitute of statistical works.

The only statistical research was performed by Akiyama \& Hara
(2000) for a time of a low solar activity. They studied the
association between flares, XPEs, and prominence eruptions in 1996
using the {\it Yohkoh}/SXT images and the H$\alpha$ reports from the
{\it Solar-Geophysical Data} (SGD). They found that 12 from 29
flares observed in the H$\alpha$ line were associated with XPEs. 16
flares were physically connected with prominence eruptions, 6 flares
were associated with both erupting phenomena, an XPE and a
prominence.

\section{Association of XPEs with prominences}

Our intention was to make the first attempt of statistical research
devoted to the association between XPEs and prominences for a time
interval longer than one year. In our studies we used a catalogue of
XPEs that resides online at http:
//www.astro.uni.wroc.pl/XPE/catalogue.html. The catalogue contains
records of 383 XPEs observed by the {\sl Yohkoh}/SXT during the full
satellite operation (1991-2001) and is corrected and upgraded
regularly. Tomczak \& Chmielewska (2012) gave a detailed description
of the catalogue as well as a statistical study of its content.

For each event from the XPEs catalogue we looked for an H$\alpha$
counterpart in a list of active prominences and filaments collected
in comprehensive reports of the SGD. In order to check a temporal
association between a prominence and an XPE, we required time of the
XPE occurrence to fall within 3-hours-interval centered around the
first observation time of the prominence. In order to check a
spatial association, we required the XPE occurrence within the same
active region as the prominence.

We obtained that only 93 from 356 XPEs (26\%) were associated with
prominences. For well-observed XPEs (so-called quality A or B in the
catalogue) this correlation is slightly better: 64 from 208 events,
i.e. 31\%.

It should however be stressed that the data taken from the SGD are
very inhomogeneous, because they come from different observatories.
There is also lack of information concerning the time periods, in
which none of the observatories monitored the Sun. These limits may
be responsible for lowering the rate of the association between XPEs
and prominences.

Therefore we compared also events from the XPEs catalogue to a list
of 186 moving prominences published by Gopalswamy {\it et al.}
(2003), They used microwave observations taken by the Nobeyama
Radioheliograph (Nakajima {\it et al.}, 1994) at 17 GHz in the years
1996-2001. From 65 limb XPEs that occurred in the years 1996-2001
during ``duty hours'' of the Nobeyama Radioheliograph (22:30-06:30
UT) only 13 events had its microwave counterpart, i.e. 20\% of
events.

To recognize subclasses of XPEs that better correlate with
prominences, we used a classification scheme developed in the
catalogue. It bases on three criteria concerning: (a) the morphology
of the XPE, (b) its kinematics, and (c) multiplicity of the
occurrence. For each criterion we distinguish two subclasses of
events: (a) collimated/loop-like; (b) confined/eruptive; (c)
single/recurrent.

In our work we used only the criteria (a) and (b), because the data
from the SGD does not inform about multiplicity of the prominence
occurrence. The morphological criterion resolves the direction of
the moving soft X-ray plasma in comparison with the local magnetic
field. In the case of the subclass 1, the direction is parallel,
i.e. along the already existing field lines; in the case of the
subclass 2 --- perpendicular, i.e. across the already existing field
lines (or strictly speaking -- together with them).

For the assignment into one of the kinematical subclasses we have
chosen a rate of the height increment above the chromosphere,
$\dot{h}$. A negative value, $\dot{h} < 0$, means the subclass 1,
the opposite case, $\dot{h} \ge 0$ means the subclass 2. XPEs from
the kinematical subclass 1 suggest the presence of magnetic or
gravitational confinement. For XPEs from the kinematical subclass 2,
an increasing velocity in the radial direction leads to
irreversible changes (eruption) of the local magnetic field. In
consequence, at least a part of the plasma escapes from the Sun.

\begin{table}[t]
\caption{The correlation with moving prominences for particular
subclasses of XPEs$^{\rm a}$}
\begin{center}
\begin{tabular}{lccc}
 \hline
 XPE subclass & Number of & Number of associated & Rate(\%)\\
 & XPEs & prominences & \\
 \hline
 \hline
collimated, confined & 84 (31)& 16 (7) & 19\% (22\%)\\
collimated, eruptive  & 30 (17)& 11 (8) & 37\% (47\%)\\
loop-like, confined  &  87 (43)& 16 (8) & 18\% (19\%)\\
loop-like, eruptive & 155 (116)& 44 (35) & 28\% (30\%)\\
total & 356 (208)& 93 (64) & 26\% (31\%)\\
\hline
\end{tabular}
\begin{list}{}{}
\item[$^{\rm a}$] Numbers in parenthesis concern well-observed XPEs
\end{list}
\label{tab1}
\end{center}
\end{table}

Table~\ref{tab1} summarizes the correlation with prominences for
particular subclasses of XPEs. As we see, the association can vary
from the average value reaching only about 20\% for confined XPEs
and almost 50\% for well-observed collimated and eruptive XPEs.

In order to compare morphology and kinematics of the associated
XPE-prominence pairs, we organized prominences using the
classification scheme developed for XPEs. Table~\ref{tab2} presents
our assignment of particular prominence subclasses applied in the
SGD to subclasses defined basing on morphology and kinematics.

\begin{table}[t]
\caption{Assignment of particular prominence subclasses (from the
SGD) to subclasses defined basing on morphology and kinematics}
\begin{center}
\begin{tabular}{cc}
 \hline
 Classification of prominences & Classification of prominences \\
 in the SGD  & in this issue \\
 \hline
 \hline
Active Dark Filament (ADF) & loop-like, confined \\
Arch Filament System (AFS) & loop-like, confined \\
Active Prominence (APR) & loop-like, confined \\
Active Surge Region (ASR) & collimated, confined \\
Bright Surge on Disk (BSD) & collimated, confined \\
Bright Surge on Limb (BSL) & collimated, confined \\
Dark Surge on Disk (DSD) & collimated, confined \\
Disappearing Filament (DSF) & loop-like, eruptive \\
Eruptive Prominence on Limb (EPL) & loop-like, eruptive \\
Loops (LPS) & loop-like, confined \\
Mound Prominence (MDP) & collimated, confined \\
Sudden Disappearing Filament (SDF) & loop-like, eruptive \\
Spray (SPY) & collimated, eruptive \\
\hline
\end{tabular}
\label{tab2}
\end{center}
\end{table}

Table~\ref{tab3} summarizes the results of this investigation. Each
row presents one subclass of XPEs established by combining both
classifying criteria, morphological and kinematical. In the first
column, the name of a subclass and the total number of XPEs in the
catalogue, for which we found associated prominences, is given. In
the next columns the counts of associated prominences are given for
particular subclasses separately. These subclasses are defined in
the same way as for XPEs (Table~\ref{tab2}). Moreover, a rate
contribution of particular subclasses is written in parenthesis.

\begin{table}[t]
\caption{Resemblance between the associated XPEs and
prominences}
\begin{center}
\begin{tabular}{lllll}
 \hline
  & \multicolumn{4}{c}{Prominences} \\
 \cline{2-5}
 XPEs & collimated, & collimated, & loop-like, & loop-like, \\
 & confined & eruptive & confined & eruptive \\
\hline \hline
 collimated, & & & & \\
 confined (16) & {\bf 6 (38\%)} & 2 (12\%)& 4 (25\%) &  4 (25\%) \\
 & & & & \\
 collimated, & & & & \\
 eruptive (11) & 8 (73\%) & {\bf 1 (9\%)} & 2 (18\%) & 0 (0\%) \\
 & & & & \\
 loop-like, & & & & \\
 confined (16) & 6 (38\%) & 1 (6\%) & {\bf 5 (31\%)} & 4 (25\%) \\
 & & & & \\
 loop-like, & & & & \\
 eruptive (44) & 18 (41\%) & 1 (2\%) & 15 (34\%) & {\bf 10 (23\%)} \\
 & & & & \\
 \hline
\end{tabular}
\label{tab3}
\end{center}
\end{table}

The bold-faced numbers situated diagonally in Table~\ref{tab3}
represent the case, in which we conclude the morphological and
kinematical similarity between the associated XPE and prominence. As
we see, only a minority of events shows this similarity (9-38\% for
particular subclasses).

To determine influence of which criterium is responsible for the
discrepancy between XPEs and prominences, we verified resemblance between
the associated XPEs and prominences for each classifying criterion separately.
Tables \ref{tab4} and \ref{tab5} present the results for the morphological and
kinematical criterion, respectively. They are organized in the same way as
Table~\ref{tab3}.

\begin{table}[t]
\caption{Morphological resemblance between the associated XPEs and
prominences}
\begin{center}
\begin{tabular}{lll}
\hline
 XPEs & \multicolumn{2}{c}{Prominences} \\ \cline{2-3}
 & collimated & loop-like \\  \hline \hline
 collimated (27)& {\bf 17 (63\%)} & 10 (37\%) \\
 & & \\
 loop-like (60)& 26 (43\%)& {\bf 34 (57\%)} \\ \hline
\end{tabular}
\label{tab4}
\end{center}
\end{table}

\begin{table}[t]
\caption{Kinematical resemblance between the associated XPEs and
prominences}
\begin{center}
\begin{tabular}{lll}
\hline
 XPEs & \multicolumn{2}{c}{Prominences} \\ \cline{2-3}
 & confined & eruptive \\  \hline \hline
 confined (32) & {\bf 21 (66\%)} & 11 (34\%) \\
 & & \\
 eruptive (55) & 43 (78\%) & {\bf 12 (22\%)} \\ \hline
\end{tabular}
\label{tab5}
\end{center}
\end{table}

As we can see, 51 from 87 (59\%) XPEs, for which we found the
associated prominences, is morphologically the same as the
prominence. For 36 events (41\%) the discrepancy exists: the
loop-like XPEs are associated with the collimated prominences or the
collimated XPEs are associated with the loop-like prominences.

Considering the kinematical criterion (Table~\ref{tab5}) we found
resemblance between the associated XPEs and prominences only for 33
from 87 (38\%) events. The discrepancy is caused mainly by the
eruptive XPEs, for which the majority (78\%) of the associated
prominences was classified as confined.

\section{Conclusions and future plans}

Our results show that XPEs and prominences are poorer correlated
than XPEs and CMEs or prominences and CMEs. Only less than 30\% of
XPEs is associated with prominences, whereas the correlation in two
last cases is about 70\%. We found subclasses of XPEs that show
higher correlation than the whole population, e.g. collimated and
eruptive XPEs, but even in this case the correlation is below 50\%
(see 2nd row of Table~\ref{tab1}). We interpret these results as a
proof that there is, in the statistical sense, only a modest
connection between hot and cool plasma motions in the corona.

Apparent discrepancy of a common occurrence in the triangle:
prominences-XPEs-CMEs (Figure~1) can be explained under assumption
that the two following subclasses of CMEs exist: (1) associated with
prominences and (2) associated with XPEs. Keeping in mind that XPEs
show almost 100\% correlation with flares, we conclude that our
results recall a traditional division of CMEs proposed by Munro {\it
et al.} (1979). The presence of two separate subclasses of CMEs,
that are correlated either with prominences or with flares, is a
subject of permanent debate. Although this division is not supported
by recent results of kinematical evolution of CMEs (e.g.
Vr$\check{\rm{s}}$nak {\it et al.}, 2005), from the theoretical
point of view it is possible that different physical mechanisms
responsible for a CME occurrence result in a different level of
association with other solar-activity phenomena (Chen, 2011)

Note that the examples in which simultaneous occurrence of an XPE
and a prominence was found, indicate also that there is only a
modest connection between hot and cool plasma motions in the corona,
since frequently there is a discrepancy between morphology and
kinematics of simultaneously occurring events (Table~\ref{tab3}).

We could not discuss a relative location of hot and cool plasma
ejections because H$\alpha$ images were unreachable for us. However,
under assumption of common vertical expansion of hot and cool
plasma, our results suggest the relative location, in which the hot
plasma is concentrated higher than the cool plasma (see
Table~\ref{tab5}: 43 examples of eruptive XPEs and the confined
prominences; 11 examples of confined XPEs and the eruptive
prominences).

To make further progress in our investigation, we are going to find
counterparts of events from the XPEs catalogue in EUV images
obtained by the {\it Transition Region And Corona Explorer} ({\it
TRACE}) satellite. Behavior of warm plasma (1-2 MK), the most
typical for the solar corona, should derive additional information
concerning interplay between hot and cool plasma. The most
interesting, well-observed events will be analyzed separately.
Unusual possibilities to make progress in a description of eruptive
phenomena in the corona are offered by the new generation of solar
orbiting observatories like {\it Hinode}, {\it Solar-Terrestrial
Relations Observatory} ({\it STEREO}), and the {\sl Solar Dynamics
Observatory} ({\sl SDO}).

\section*{Acknowledgements}
The {\sl Yohkoh} satellite is a project of the Institute of Space
and Astronautical Science of Japan. We acknowledge solar data
collected and distributed by U. S. National Geophysical Data Center.


\section*{References}
\begin{itemize}
\small
\itemsep -2pt
\itemindent -20pt
\item[] Akiyama, S., Hara, H.: 2000, The Last Total Solar Eclipse of
the Millennium in Turkey (W. Livingston, A. $\ddot{\rm{O}}$zg$\ddot{\rm{u}}$c,
eds.), {\it ASP Conference Series}, {\bf 205}, 137.
\item[] Chen, P.\,F.: 2011, {\it Living Rev. Solar Phys.}, {\bf 8}, 1.
\item[] de Jager, C.: 1959, {\it Handbuch der Physik}, {\bf 52}, 80.
\item[] Gilbert, H.\,R., Holzer, T.\,E., Burkepile, J.\,T., Hundhausen,
A.\,J.: 2000, {\it Astrophys.\,J.}, {\bf 537}, 503.
\item[] Gopalswamy, N., Kundu, M.\,R., Manoharan, P.\,K., Raoult,
A., Nitta, N., Zarka, P.: 1997, {\it Astrophys.\,J.}, {\bf 486},
1036.
\item[] Gopalswamy, N., Shimojo, M., Lu, W., Yashiro, S., Shibasaki,
K., Howard, R.\,A.: 2003, {\it Astrophys.\,J.}, {\bf 586}, 562.
\item[] Gopalswamy, N., Yashiro, S., Micha{\l }ek, G., Stenborg, G.,
Vourlidas, A., Freeland, S., Howard, R.: 2009, {\it Earth Moon
Planet}, {\bf 104}, 295.
\item[] Kim,\,Y.-H., Moon,\,Y.-J., Cho,\,K.-S., Kim, K.-S., Park, Y.\,D.:
2005, {\it Astrophys.\,J.}, {\bf 622}, 1240.
\item[] Kim,\,Y.-H., Bong, S.-Ch., Park, Y.-D., Cho,\,K.-S.,
Moon,\,Y.-J.: 2009, {\it Astrophys.\,J.}, {\bf 705}, 1721.
\item[] Mari$\check{\rm{c}}$i$\acute{\rm{c}}$, D., Vr$\check{\rm{s}}$nak, B.,
Stanger, A.\,L., Veronig, A.: 2004, {\it Solar Phys.}, {\bf 225},
337.
\item[] Munro, R.\,H., Gosling, J.\,T., Hildner, E., MacQueen, R.\,M.,
Poland, A.\,I., Ross, C.\,L.: 1979, {\it Solar Phys.}, {\bf 61},
201.
\item[] Nakajima, H., {\it et al.}: 1994, {\it Proc. IEEE}, {\bf 82}, 705.
\item[] Ohyama, M., Shibata, K.: 2008, {\it Publ.\,Astron.\,Soc.\,Japan},
{\bf 60}, 85.
\item[] Secchi, A.: 1875, {\it Le Soleil}, Gauthier-Villars, Paris.
\item[] Shibata,\,K., Masuda,\,S., Shimojo,\,M., Hara, H., Yokoyama, T.,
Tsuneta, S., Kosugi, T., Ogawara, Y.: 1995, {\it
Astrophys.\,J.\,Lett.}, {\bf 451}, L83.
\item[] Tandberg-Hanssen, E.: 1995, {\it The Nature of Solar Prominences},
Kluwer, Dordrecht.
\item[] Tomczak, M., Chmielewska, E..: 2012, {\it
Astrophys.\,J.\,Suppl.}, {\bf 199}, 10.
\item[] Tousey, R.: 1973, {\it Space\,Res.}, {\bf XIII}, 713.
\item[] Tsuneta,\,S., {\it et al.}: 1991, {\it Solar Phys.},
{\bf 136}, 37.
\item[] Vr$\check{\rm{s}}$nak, B., Mari$\check{\rm{c}}$i$\acute{\rm{c}}$, D.,
Stanger, A.\,L., Veronig, A.: 2004, {\it Solar Phys.}, {\bf 225},
355.
\item[] Vr$\check{\rm{s}}$nak, B., Sudar, D., Ru$\check{\rm{z}}$djak, D.:
 2005, {\it Astron.\,Astrophys.}, {\bf 435}, 1149.
\end{itemize}

\bibliographystyle{ceab}
\bibliography{sample}

\end{document}